\documentclass[twocolumn,showpacs,preprintnumbers,amsmath,amssymb]{revtex4}

\usepackage{graphicx}
\usepackage{dcolumn}
\usepackage{bm}
\usepackage{mathrsfs}

\begin{document}

\preprint{APS/123-QED}

\title{Characterization of $^7$H Nuclear System}

\author{M. Caama\~no$^{1,2}$\footnote{Electronic address: {\tt mfresco@usc.es}}, D. Cortina-Gil$^1$, W. Mittig$^2$, H. Savajols$^2$, M. Chartier$^3$, C. E. Demonchy$^2$, B. Fern\'andez$^3$, \mbox{M. B. \surname{G\'omez Hornillos}$^3$}, A. Gillibert$^4$, B. Jurado$^2$, O. Kiselev$^{5,6}$, R. Lemmon$^7$,  A. Obertelli$^4$, F. Rejmund$^2$, M. Rejmund$^2$, P. Roussel-Chomaz$^2$, R. Wolski$^8$}
\affiliation{$^1$Dept. of Particle Physics, Universidade de Santiago de Compostela,
E-15782, Santiago de Compostela, Spain}
\affiliation{$^2$GANIL CEA/DSM-CNRS/IN2P3, BP 55027, 14076 Caen Cedex 05, France}
\affiliation{$^3$Dept. of Physics, University of Liverpool,
Oliver Loge Laboratory, L69 7ZE, United Kingdom}
\affiliation{$^4$CEA/DSM/DAPNIA,
CEA Saclay, 91191 Gif sur Yvette Cedex, France}
\affiliation{$^5$Inst. of Nuclear Chemistry, University of Mainz, 55128 Mainz, Germany}
\affiliation{$^6$PNPI, Gatchina 188300, Russia}
\affiliation{$^7$CCLRC Daresbury Laboratory,
Warrington, Cheshire, WA4 4AD, United Kingdom}
\affiliation{$^8$Henryk Niewodnicza\'nski Institute of Nuclear Physics,
ul. Radzikowskiego 152 31-342 Krakow, Poland}

\date{\today}

\begin{abstract}
The $^7$H resonance was produced via one-proton transfer reaction with a $^8$He beam at \mbox{15.4A MeV} and a $^{12}$C gas target
. The experimental setup was based on the active-target MAYA which allowed a complete reconstruction of the reaction kinematics. The characterization of the identified $^7$H events resulted in a resonance energy of 0.57$^{+0.42}_{-0.21}$ MeV above the $^3$H+4n threshold and a resonance width of 0.09$^{+0.94}_{-0.06}$ MeV. 
\end{abstract}

\maketitle


A major goal in nuclear physics is to understand how nuclear stability and structure arise from the underlying interaction between individual nucleons. 
Systematic measurements of nuclei far from stability are a valuable tool to test the present models, which are mainly based on properties of stable nuclear matter, and check the validity of their predictions extended to exotic nuclei. Recent developments in the production of radioactive beams, as ISOL and In-Flight techniques, bring new opportunities to study these nuclei
. The study of resonances beyond the drip lines is relatively accessible for neutron-rich light nuclei, such as hydrogen and helium, where the drip line is reached with the addition of only a few neutrons to the stable isotopes.

Even if the search for hydrogen isotopes heavier than tritium started more than 30 years ago\cite{4}, the map of the super-heavy hydrogen isotopes is far from being complete
. Whereas experiments have confirmed the existence of $^4$H, $^5$H and $^6$H as resonances\cite{5,6,7,8,9}, their fundamental properties are not unambiguously determined. 
This work concentrates on the search for the $^7$H resonance, the nuclear system with the most extreme neutron to proton ratio presently reached, as a further step in the systematic study of hydrogen resonances. 

From the theoretical point of view, the case of light nuclei is particularly interesting because the number of nucleons involved places these species in between two scenarios. They can be seen either as few-particle systems that can be studied directly from the nucleon-nucleon interaction\cite{14,15,16} or as many-body systems where inner structure dominates\cite{11,12,13}. A common source of uncertainty in all these approaches is the use of effective nucleon-nucleon interactions, which is particularly delicate in the case of the neutron-neutron interaction due to the lack of accurate descriptions.

Recent results from some of these approaches have predicted the existence of the $^7$H resonance with a resonance energy above the $^3$H+4n mass varying from around 1 MeV in a {\it Hyperspherical Basis} approach\cite{14} up to 7 MeV in an {\it Antisymmetrized Molecular Dynamics} calculation\cite{16}. In parallel, experimental studies performed by Korsheninnikov {\it et al.}\cite{17} 
show a sharp increase in the p($^8$He,pp) channel close to the $^3$H+4n disintegration threshold, interpreted as a first tentative evidence of the existence of $^7$H as a low lying resonance. Our work is a major step in this direction and represents a clear experimental proof of the existence of $^7$H as a nuclear system along its characterisation as a resonance.

\begin{figure} [t]
\begin{center}
\includegraphics[width=8cm]{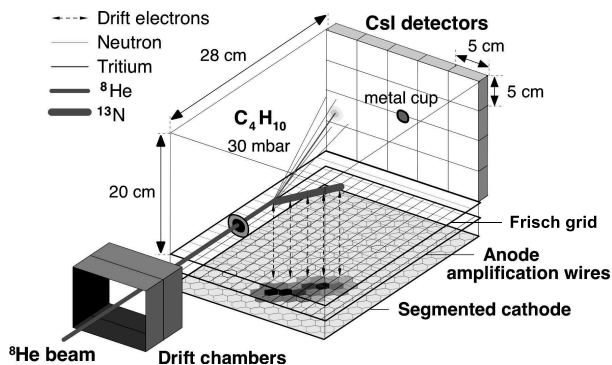}
\vspace{0.5cm}
\caption{\label{maya} Experimental setup. Two drift chambers placed before the MAYA detector monitor the beam particles. The reactions occur in the volume filled with C$_4$H$_{10}$ gas. The amplification zone contains a Frisch grid, the anode wires and the segmented ca\-tho\-de. A 5$\times$5 matrix of CsI detectors is placed at the back side for detecting light particles. A small metal cup is used to stop the beam.}
\end{center}
\end{figure}


The experiment was performed at GANIL (France) using the SPIRAL facility based on the Isotope Se\-pa\-ra\-tion On Line (ISOL) technique\cite{antonio} of beam production. A secondary beam of $^8$He at 15.4A MeV, with an intensity of $\sim$10$^4$ pps, was produced using a primary $^{13}$C beam on a thick $^{12}$C target. The $^7$H system was then studied via the $^{12}$C($^{8}$He,$^7$H $\rightarrow$ $^3$H+4n)$^{13}$N transfer reaction.

The experimental setup used in the present experiment detected the charged particles involved in the reaction using the active-target MAYA\cite{18}, which is specially well-suited for detecting reaction products in a very low energy kinematic domain. This detector is a Time-Charge Projection Chamber where the detection gas plays also the role of reaction target. 
The beam particles and the reaction products ionize the gas along their paths. The electrons released in the ionization process drift toward the amplification area where they are accelerated around a plane of amplification wires after traversing a Frisch grid. The accelerated electrons ionize again the surrounding gas inducing a mirror charge in the pads of a segmented cathode placed below the wires. Measurements of the drift time and the charge induced on the segmented cathode enable a complete 3-dimensional tracking of those reaction products that lose enough energy to be detected. A segmented wall of twenty cesium-iodide (CsI) crystals placed at forward angles detects those particles that do not stop inside the gas volume. The detection of a charge particle in any CsI detector was used for triggering the acquisition during the experiment. Two drift chambers located before MAYA are used as beam monitors
. The non-reacting projectiles are stopped in a small metal cup at the end of MAYA. Figure \ref{maya} shows a schematic view of the experimental setup. 

\begin{figure} [b]
\begin{center}
\includegraphics[width=7cm]{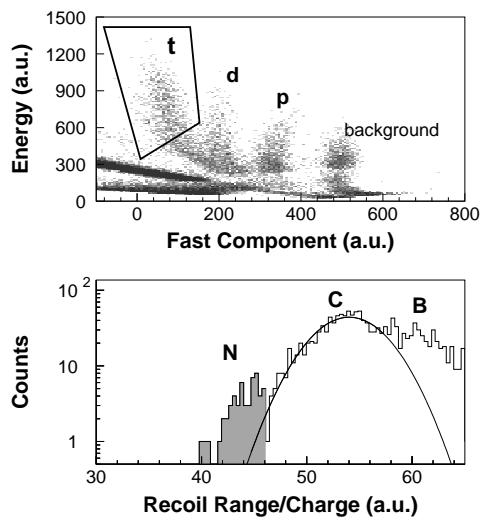}
\caption{\label{csi} Tritium and nitrogen identification. Upper panel: the selection of tritium among other hydrogen isotopes in the CsI detectors is shown in energy vs. fast component coordinates. High energy tritons are selected with the solid line. Lower panel: the selection of nitrogen among other recoil species is shown in a spectrum of range-charge ratio. The carbon isotopes region is fitted to a Gaussian shape for reference. The identified nitrogen appear as full grey histogram.}
\end{center}
\end{figure}

In a typical event where $^7$H is produced, a $^8$He projectile enters in the detector and transfers one proton to the nucleus of a $^{12}$C atom of the gas, C$_4$H$_{10}$ at 30 mbar, which corresponds to a target thickness of $3.2\cdot 10^{19}$ $^{12}$C/cm$^{2}$. The scattered $^7$H decays immediately into $^3$H and four neutrons. The first step in the selection of the $^{12}$C($^8$He,$^7$H $\rightarrow$ $^3$H+4n)$^{13}$N \mbox{channel} consists in the coincident identification of the charged reaction products, tritium and nitrogen. The triton is stopped in the segmented CsI wall and identified via the relation between the total energy and the fast component of the CsI signal output, which is sensitive to the mass and charge of the particle \cite{knoll} (Figure \ref{csi}). The nitrogen recoil, with a total energy between 3 and 15 MeV, corresponding to ranges between 40 and 160 mm, is stopped inside the detector. The range and angle are measured using the charge \mbox{image} projected on the segmented cathode, with typical uncertainties of $\pm$2 mm and $\pm$5 deg respectively. The identification of the recoil is done by means of the relation between the measured range and the deposited charge, which is a function of the total energy when the recoil is completely stopped inside the gas. The nitrogen total energy is then calculated from the measured range using the available code SRIM \cite{srim}. Figure \ref{csi} shows the selection of nitrogen among other recoil species by means of their different range over charge ratios.
The largest peak corresponds to carbon isotopes mainly coming from elastic reactions. Higher range over charge ratios are populated with isotopes with lower charges, such as boron isotopes produced in $^{12}$C($^8$He,$^x$Li)$^{20-x}$B reactions. The right peak is populated with charges greater than carbon, which in the present case can only correspond to nitrogen isotopes when a charge particle detected in coincidence in the CsI wall is required. The different 1{\it p}-x{\it n} transfer reaction channels producing $^3$H and a nitrogen recoil are separated afterwards by their different kinematics. Contributions from other reaction channels, such as fusion-evaporation, are eliminated with the coincident detection of a single recoil, identified as nitrogen, and a single scattered particle in the CsI wall, identified as a triton with relatively high energy.


The $^{12}$C($^8$He,$^7$H)$^{13}$N one-proton transfer is a binary reaction with two particles in the final state. Conservation of energy and momentum allows to reconstruct the reaction kinematics from the information of only one of the reaction products. In the present work, the reconstruction is done with the nitrogen recoil angles and energies measured with the 3-dimensional tracking of MAYA. The kinematic information can be reduced to the excitation energy of the $^7$H system applying a missing mass calculation. The excitation energy is then defined as the difference between the calculated mass of the $^7$H system with respect to the reference $^3$H+4n sub-system mass.

\begin{figure} [b]
\begin{center}
\includegraphics[width=7cm]{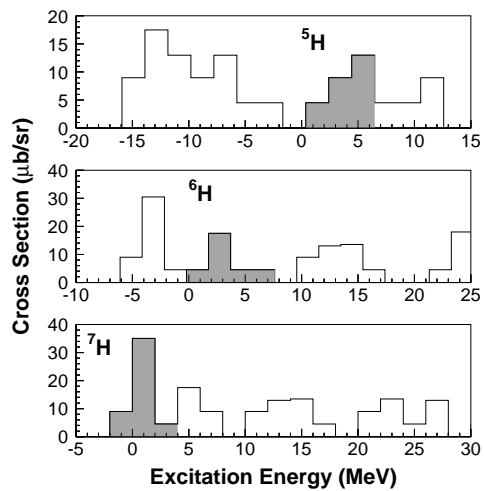}
\caption{\label{cine} Excitation energy distributions calculated under the assumptions of $^5$H (upper panel), $^6$H (middle panel) and $^7$H (lower panel) production channels. The grey histograms in the $^5$H and $^6$H channels correspond to those events laying in the regions where the resonances were already observed in previous experiments. The grey histogram in the $^7$H corresponds to those events identified as $^7$H production. See text for details.
}
\end{center}
\end{figure}

The identification of the $^7$H events is done after the identification of the events corresponding to other reaction channels with a $^3$H and a nitrogen products, such as $^6$H and $^5$H. Upper panel in Figure \ref{cine} shows the excitation energy distribution corresponding to $^5$H production. This is calculated assuming that the detected nitrogen is $^{15}$N and the sub-system mass is $^3$H+2n. Those events marked as a grey histogram lay on the region where $^5$H is expected according to previous experiments \cite{7,8}, and they are associated to the $^5$H channel. Middle panel shows the excitation energy distribution corresponding to $^6$H production. The calculation was done assuming the detected nitrogen as $^{14}$N and a $^3$H+3n subsystem mass. The events in the peak marked as a grey histogram are different from those associated with $^5$H and lay on the region where $^6$H was observed in previous experiments \cite{9}. These events correspond to the $^6$H channel. Finally, the lower panel shows the excitation energy distribution assuming the detected nitrogen as $^{13}$N and the sub-system mass as $^3$H+4n. The events laying on the peak marked in grey around the $^3$H+4n disintegration threshold are different from those previously associated to $^5$H and $^6$H channels. In addition, other reactions produced in the present experimental setup, such as fusion-evaporation, are estimated to populate the kinematic region of the $^7$H with less than one count after the $^3$H+nitrogen selection. Estimations on the six-body $^{13}$N+$^3$H+4n phase-space of the $^7$H channel, and phase-space associated with $^6$H and $^5$H channels result in a background contribution which begins to be appreciable around -10 MeV below the $^3$H+4n threshold. 
Under these considerations those events located in the marked region are identified as $^7$H production. The peaked distribution is a signature of a well-defined state and represents a background free confirmation of the production of the $^7$H resonance.


Seven events corresponding to the $^7$H resonance were identified in the present analysis
. The cross section of the $^7$H production was determined as the number of detected events normalized to the number of incident projectiles and target nuclei. This calculation is corrected by the efficiency of the detection system. A mean differential cross section of d$\sigma$/d$\Omega = 40.1^{+58.0}_{-30.6} ~\mu$b/sr, was obtained within the angular coverage of MAYA, calculated as \mbox{9.7 - 48.2 deg} in the centre of mass frame. 

\begin{figure} [t]
\begin{center}
\includegraphics[width=7cm]{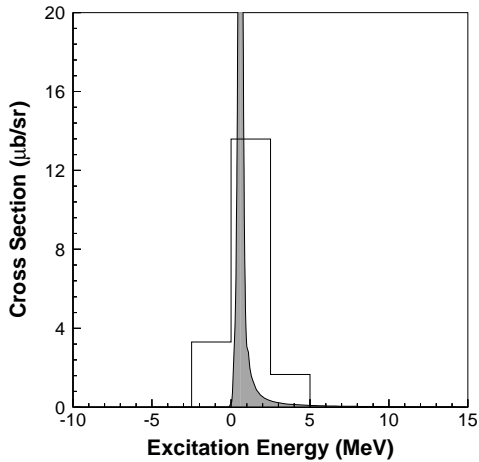}
\caption{\label{bw} Excitation energy distribution for the identified $^7$H events. The solid function is the Breit-Wigner distribution resulting from the fit to the experimental events, represented with the empty histogram. The 2.5 MeV binning corresponds to the mean energy uncertainty.}
\end{center}
\end{figure}

The peak in excitation energy corresponding to the production of $^7$H (Figure \ref{bw}) is described in this work with a modified Breit-Wigner distribution\cite{19}: 

\begin{equation}
\sigma_{BW}=\sigma_0\frac{\Gamma\sqrt{\frac{E^{exc}}{E_R}}}{(E^{exc}-E_R)^2-\frac{\Gamma^2}{4}\frac{E^{exc}}{E_R}}
\end{equation}

where the production cross section, $\sigma_{BW}$, depends on the excitation e\-ner\-gy, $E^{exc}$, through the resonance ener\-gy, $E_R$, and width, $\Gamma$. The formula includes a modification factor, $\sqrt{\frac{E^{exc}}{E_R}}$, to take into account energy dependence of the system barrier. The factor $\sigma_0$ is determined with the normalization to the total cross section.

The Breit-Wigner function is fitted to the experimental values of the excitation energy using a multi-parametric Maximum Likelihood procedure, which is especially suited to low statistics samples. 


The Likelihood is calculated 
associating to each event a Gaussian function centered in the measured excitation energy and with a variance equal to its calculated uncertainty. The width of the experimental distribution can be described as a mathematical convolution of the Breit-Wigner distribution and a function associated with the experimental uncertainty of the data set, which is around 2.5 MeV. The Maximum Likelihood procedure disentangles both contributions to the final result.

The fitted parameters result in a width of \mbox{$\Gamma = 0.09^{+0.94}_{-0.06}$ MeV}, and a resonance energy of \mbox{$E_R = 0.57^{+0.42}_{-0.21}$ MeV} above the threshold of the $^3$H+4n sub-system. The results are shown in Table \ref{tab1}. In Figure \ref{bw} with the fitted Breit-Wigner distribution is displayed over the excitation energy distribution of the $^7$H detected events.

\begin{table}
\caption{\label{tab1}$^7$H resonance characterization parameters.}
\begin{ruledtabular}
\begin{tabular}{cccc}
 $E_R$ (MeV) & $\Gamma$ (MeV) & $\sigma_0$ ($\mu$b/sr) &
 d$\sigma_{BW}$/d$\Omega$ ($\mu$b/sr) \\
\hline\\
$0.57^{+0.42}_{-0.21}$ & $0.09^{+0.94}_{-0.06}$ & $6.4^{+9.0}_{-4.9}$ & $40.1^{+58.0}_{-30.6}$  \\
~&~&~&~\\
\end{tabular}
\end{ruledtabular}
\end{table}


The previous experimental observation of Korsheninnikov {\it et al.}\cite{17}, where a sharp increase of the cross section appeared close to the $^3$H+4n threshold, is in qualitative agreement with the resonance energy evaluated in this work. Regarding theoretical descriptions, calculations based on a {\it Hyperspherical Basis} of the shell model are closest to the present work\cite{14,15}, even though the predicted resonance energies range between 1 MeV \cite{14} and 3 MeV \cite{15}. In any case, it is difficult to conclude that this is the appropriate description due to the lack of predictions for $^7$H from other approaches. Estimations of the width of the resonance were also done in a work of Golovkov {\it et al.}\cite{20} resulting in a theoretical width around three orders of magnitude lower than the present work. The resonance width is related to the decay rate and it contains information about the decay mechanism. The resulting narrow width extracted in this work may be a hint of a fast and unique four-neutron decay\cite{17,21}. Future studies of the $^7$H nuclear state may clarify this point, also interesting for the question about the existence of a bound or unbound tetra-neutron ($^4$n) state\cite{22,23}.


The present results constitute a major step forward in the existence of the most exotic nuclear systems ever found, showing that nuclear matter with N/Z up to 6 can still exist. They 
provide essential input for de\-ve\-loping theoretical descriptions and improving in general our understanding of nuclear matter. 

\begin{acknowledgments}
This work was supported by grants from the Spanish Ministry of Education and Science, Conseller\'\i a de Innovaci\'on e Industria da Xunta de Galicia, the CICYT-IN2P3 cooperation, and the Marie Curie Fellowship support (M.Caama\~no). The authors wish to thank deeply Dr. A. Navin for valuable discussions during the writing of this article.\\
\end{acknowledgments}

\bibliography{caamano}

\begin{thebibliography}{22}
\expandafter\ifx\csname natexlab\endcsname\relax\def\natexlab#1{#1}\fi
\expandafter\ifx\csname bibnamefont\endcsname\relax
  \def\bibnamefont#1{#1}\fi
\expandafter\ifx\csname bibfnamefont\endcsname\relax
  \def\bibfnamefont#1{#1}\fi
\expandafter\ifx\csname citenamefont\endcsname\relax
  \def\citenamefont#1{#1}\fi
\expandafter\ifx\csname url\endcsname\relax
  \def\url#1{\texttt{#1}}\fi
\expandafter\ifx\csname urlprefix\endcsname\relax\def\urlprefix{URL }\fi
\providecommand{\bibinfo}[2]{#2}
\providecommand{\eprint}[2][]{\url{#2}}

\bibitem[{\citenamefont{Adelberg et~al.}(1967)}]{4}
\bibinfo{author}{\bibfnamefont{E.~G.} \bibnamefont{Adelberg}}
  \bibnamefont{et~al.}, \bibinfo{journal}{Phys. Lett. B}
  \textbf{\bibinfo{volume}{25}}, \bibinfo{pages}{595} (\bibinfo{year}{1967}).

\bibitem[{\citenamefont{Belozyorov et~al.}(1986)}]{5}
\bibinfo{author}{\bibfnamefont{A.~V.} \bibnamefont{Belozyorov}}
  \bibnamefont{et~al.}, \bibinfo{journal}{Nucl. Phys. A}
  \textbf{\bibinfo{volume}{460}}, \bibinfo{pages}{352} (\bibinfo{year}{1986}).

\bibitem[{\citenamefont{Sidorchuk et~al.}(2004)}]{6}
\bibinfo{author}{\bibfnamefont{S.~I.} \bibnamefont{Sidorchuk}}
  \bibnamefont{et~al.}, \bibinfo{journal}{Phys. Lett. B}
  \textbf{\bibinfo{volume}{594}}, \bibinfo{pages}{54} (\bibinfo{year}{2004}).

\bibitem[{\citenamefont{Meister et~al.}(2003)}]{7}
\bibinfo{author}{\bibfnamefont{M.}~\bibnamefont{Meister}} \bibnamefont{et~al.},
  \bibinfo{journal}{Phys.\ Rev. Lett.} \textbf{\bibinfo{volume}{91}},
  \bibinfo{pages}{162504} (\bibinfo{year}{2003}).

\bibitem[{\citenamefont{Korsheninnikov
  et~al.}(2001)\citenamefont{Korsheninnikov, Golovkov, and Tanihata}}]{8}
\bibinfo{author}{\bibfnamefont{A.~A.} \bibnamefont{Korsheninnikov}},
  \bibinfo{author}{\bibfnamefont{M.~S.} \bibnamefont{Golovkov}},
  \bibnamefont{and} \bibinfo{author}{\bibfnamefont{I.}~\bibnamefont{Tanihata}},
  \bibinfo{journal}{Phys.\ Rev. Lett.} \textbf{\bibinfo{volume}{87}},
  \bibinfo{pages}{092501} (\bibinfo{year}{2001}).

\bibitem[{\citenamefont{Aleksandrov et~al.}(1984)}]{9}
\bibinfo{author}{\bibfnamefont{D.}~\bibnamefont{Aleksandrov}}
  \bibnamefont{et~al.}, \bibinfo{journal}{Sov. J. Nucl. Phys.}
  \textbf{\bibinfo{volume}{39}}, \bibinfo{pages}{323} (\bibinfo{year}{1984}).

\bibitem[{\citenamefont{Timofeyuk}(2002)}]{14}
\bibinfo{author}{\bibfnamefont{N.~K.} \bibnamefont{Timofeyuk}},
  \bibinfo{journal}{Phys.\ Rev. C} \textbf{\bibinfo{volume}{65}},
  \bibinfo{pages}{064306} (\bibinfo{year}{2002}).

\bibitem[{\citenamefont{Timofeyuk}(2004)}]{15}
\bibinfo{author}{\bibfnamefont{N.~K.} \bibnamefont{Timofeyuk}},
  \bibinfo{journal}{Phys.\ Rev. C} \textbf{\bibinfo{volume}{69}},
  \bibinfo{pages}{034336} (\bibinfo{year}{2004}).

\bibitem[{\citenamefont{Aoyama and Itagaki}(2004)}]{16}
\bibinfo{author}{\bibfnamefont{S.}~\bibnamefont{Aoyama}} \bibnamefont{and}
  \bibinfo{author}{\bibfnamefont{N.}~\bibnamefont{Itagaki}},
  \bibinfo{journal}{Nucl. Phys. A} \textbf{\bibinfo{volume}{738}},
  \bibinfo{pages}{362} (\bibinfo{year}{2004}).

\bibitem[{\citenamefont{Blanchon et~al.}(2004)\citenamefont{Blanchon,
  Bonaccorso, and Vinh~Mau}}]{11}
\bibinfo{author}{\bibfnamefont{G.}~\bibnamefont{Blanchon}},
  \bibinfo{author}{\bibfnamefont{A.}~\bibnamefont{Bonaccorso}},
  \bibnamefont{and} \bibinfo{author}{\bibfnamefont{N.}~\bibnamefont{Vinh~Mau}},
  \bibinfo{journal}{Nucl. Phys. A} \textbf{\bibinfo{volume}{739}},
  \bibinfo{pages}{259} (\bibinfo{year}{2004}).

\bibitem[{\citenamefont{Descouvemont and Kharbach}(2001)}]{12}
\bibinfo{author}{\bibfnamefont{P.}~\bibnamefont{Descouvemont}}
  \bibnamefont{and} \bibinfo{author}{\bibfnamefont{A.}~\bibnamefont{Kharbach}},
  \bibinfo{journal}{Phys. Rev. C} \textbf{\bibinfo{volume}{63}},
  \bibinfo{pages}{027001} (\bibinfo{year}{2001}).

\bibitem[{\citenamefont{Arai}(2003)}]{13}
\bibinfo{author}{\bibfnamefont{K.}~\bibnamefont{Arai}},
  \bibinfo{journal}{Phys.\ Rev. C} \textbf{\bibinfo{volume}{68}},
  \bibinfo{pages}{034303} (\bibinfo{year}{2003}).

\bibitem[{\citenamefont{Korsheninnikov et~al.}(2003)}]{17}
\bibinfo{author}{\bibfnamefont{A.~A.} \bibnamefont{Korsheninnikov}}
  \bibnamefont{et~al.}, \bibinfo{journal}{Phys.\ Rev. Lett.}
  \textbf{\bibinfo{volume}{90}}, \bibinfo{pages}{082501}
  (\bibinfo{year}{2003}).

\bibitem[{\citenamefont{Villari et~al.}(1995)}]{antonio}
\bibinfo{author}{\bibfnamefont{A.~C.~C.} \bibnamefont{Villari}}
  \bibnamefont{et~al.}, \bibinfo{journal}{Nucl. Phys. A}
  \textbf{\bibinfo{volume}{588}}, \bibinfo{pages}{267c} (\bibinfo{year}{1995}).

\bibitem[{\citenamefont{Mittig et~al.}(2003)}]{18}
\bibinfo{author}{\bibfnamefont{W.}~\bibnamefont{Mittig}} \bibnamefont{et~al.},
  \bibinfo{journal}{Nucl. Phys. A} \textbf{\bibinfo{volume}{722}},
  \bibinfo{pages}{10c} (\bibinfo{year}{2003}).

\bibitem[{\citenamefont{Knoll}(1989)}]{knoll}
\bibinfo{author}{\bibfnamefont{G.~F.} \bibnamefont{Knoll}},
  \emph{\bibinfo{title}{Radiation Detection and Measurement}}
  (\bibinfo{publisher}{J. Wiley and sons, Inc.}, \bibinfo{year}{1989}).

\bibitem[{\citenamefont{Ziegler}(2005)}]{srim}
\bibinfo{author}{\bibfnamefont{J.~F.} \bibnamefont{Ziegler}}
  (\bibinfo{year}{2005}), \eprint{http://www.srim.org}.

\bibitem[{\citenamefont{Breit and Wigner}(1936)}]{19}
\bibinfo{author}{\bibfnamefont{G.}~\bibnamefont{Breit}} \bibnamefont{and}
  \bibinfo{author}{\bibfnamefont{E.}~\bibnamefont{Wigner}},
  \bibinfo{journal}{Phys.\ Rev.} \textbf{\bibinfo{volume}{49}},
  \bibinfo{pages}{519} (\bibinfo{year}{1936}).

\bibitem[{\citenamefont{Golovkov}(2004)}]{20}
\bibinfo{author}{\bibfnamefont{M.~S.} \bibnamefont{Golovkov}},
  \bibinfo{journal}{Phys. Lett. B} \textbf{\bibinfo{volume}{588}},
  \bibinfo{pages}{163} (\bibinfo{year}{2004}).

\bibitem[{\citenamefont{Korsheninnikov}(2005)}]{21}
\bibinfo{author}{\bibfnamefont{A.~A.} \bibnamefont{Korsheninnikov}},
  \bibinfo{journal}{Nucl. Phys. A} \textbf{\bibinfo{volume}{751}},
  \bibinfo{pages}{501c} (\bibinfo{year}{2005}).

\bibitem[{\citenamefont{Marqu\'es et~al.}(2002)}]{22}
\bibinfo{author}{\bibfnamefont{F.~M.} \bibnamefont{Marqu\'es}}
  \bibnamefont{et~al.}, \bibinfo{journal}{Phys.\ Rev. C}
  \textbf{\bibinfo{volume}{65}}, \bibinfo{pages}{044006}
  (\bibinfo{year}{2002}).

\bibitem[{\citenamefont{Pieper}(2003)}]{23}
\bibinfo{author}{\bibfnamefont{S.~C.} \bibnamefont{Pieper}},
  \bibinfo{journal}{Phys. Rev. Lett.} \textbf{\bibinfo{volume}{90}},
  \bibinfo{pages}{252501} (\bibinfo{year}{2003}).

\end{thebibliography}

\end{document}